\documentclass[preprint,showpacs,preprintnumbers,amsmath,amssymb]{revtex4}
\usepackage{graphicx}
\usepackage{dcolumn}
\usepackage{bm}
\usepackage{epsfig}
\def\trade{{\bigcirc}\!\!\!\!\!\mbox{{\tiny R}}}
\def\mathmath{{\it Mathematica}$^{\trade}$\,\,}
\advance\voffset by +25pt
\begin{document}

\title[Parity restricted potentials]{New identities from quantum-mechanical sum rules\\
of parity-related potentials}

\author{O. A. Ayorinde} \email{seayorinde@davidson.edu}
\author{K. Chisholm} \email{kechisholm@davidson.edu}
\author{M. Belloni} \email{mabelloni@davidson.edu}
\affiliation{%
Physics Department \\
Davidson College \\
Davidson, NC 28035 USA \\
}

\author{R. W. Robinett} \email{rick@phys.psu.edu}
\affiliation{%
Department of Physics\\
The Pennsylvania State University\\
University Park, PA 16802 USA \\
}

\begin{abstract}
We apply quantum-mechanical sum rules
to pairs of one-dimensional systems defined by
potential energy functions related by parity.
Specifically, we consider symmetric potentials, $V(x) = V(-x)$, and
their \textit{parity-restricted} partners, ones with $V(x)$ but defined only
on the positive half-line. We extend recent
discussions of sum rules for the quantum bouncer by considering the
\textit{parity-extended} version of this problem, defined by the
symmetric linear potential, $V(z) = F|z|$, and
find new classes of constraints on the zeros of the Airy
function, $Ai(\zeta)$, and its derivative, $Ai'(\zeta)$.
We also consider the parity-restricted version of the
harmonic oscillator and find completely new classes of
mathematical relations, unrelated to those of the ordinary oscillator problem.
These two soluble quantum-mechanical systems defined by
power-law potentials provide examples of how the form of the potential
(both parity and continuity properties) affect the convergence
of quantum-mechanical sum rules. We also discuss semi-classical predictions
for expectation values and the Stark effect for these systems.

\end{abstract}


\maketitle

\section{Introduction}

Sum rules, constraints on energy-difference weighted combinations of
on- and off-diagonal matrix elements, have been an important theoretical construct in
quantum mechanics since its earliest days. For example, the
Thomas-Reiche-Kuhn (TRK) \cite{trk_sum_rule} sum rule was one of the earliest
quantitative checks on the oscillator strengths of atomic transitions and
an important confirmation of the applicability of quantum mechanics to
atomic systems. Their use continues today, for example, with QCD sum
rules used to probe the masses of both light and heavy quarks in
hadronic systems. For a list of recent applications of sum rules,
see Ref.~\cite{belloni_robinett_sum_rules}.

The study of sum rules and how such constraints are  realized \cite{belloni_robinett_sum_rules}
can provide useful examples of a variety of mathematical techniques utilized for their confirmation.
In addition, sum rules can be used to generate new mathematical constraints
\cite{mavromatis} on quantities involving infinite sums.
For example, in exactly that context, two of us  have recently
shown how it is possible to systematically derive
new constraints on the zeros, ($-\zeta_n$), of the Airy function, $Ai(\zeta)$,
of the form $S_p(n) = \sum_{k \neq n} (\zeta_k - \zeta_n)^{-p}$ for
natural values of
$p \geq 2$ using quantum-mechanical sum rules applied to the so-called `quantum
bouncer' problem \cite{belloni_robinett_airy_zeros}.
Defined by the potential
\begin{equation}
V(z) =
\left\{
\begin{array}{cc}
Fz & \mbox{for $z>0$} \\
\infty &\mbox{for $z<0$}
\end{array}
\right.
\, ,
\label{bouncer_potential}
\end{equation}
this quantum-mechanical problem has been applied to recent novel
experiments, ranging from the bound states of neutrons in
the Earth's gravitational field \cite{neutron_bound_states}
to optical analogs of gravitational wave packets \cite{photon_bouncing_ball}
which demonstrate the presence of predicted \cite{bouncing_ball}
wave packet revivals.

In our earlier study, a number  of the most well-known sum rules
did not converge
due to the lack of continuity of the bouncer potential at the origin.
In this note, we extend the results of
Ref.~\cite{belloni_robinett_airy_zeros} to discuss mathematical
constraints which arise from the study of quantum-mechanical sum rules
in a closely related system, namely the symmetric linear potential,
defined by
\begin{equation}
V(z) = F|z|
\label{symmetric_linear_potential}
\, .
\end{equation}
This is an example of a \textit{parity-extended} potential, namely taking one that is
defined only for $z>0$, and extending it in a symmetric way over the
real axis.
In this case, we find new relationships involving  \emph{both} the zeros
of $Ai(\zeta)$ \emph{and} its derivative $Ai'(\zeta)$, namely $-\zeta_n$ and
$-\eta_n$, respectively. In Sec.~\ref{sec:constraints} we also explore the 
convergence properties of sum rules in this system, compared to the quantum
bouncer, because of the improved mathematical behavior of the
potential energy function in this case.

For comparison, we also consider in Sec.~\ref{sec:half_sho}
a \textit{parity-restricted} version of the most well-studied system in
quantum mechanics, namely the harmonic oscillator. Specifically, we
examine the structure of sum rules in the `half'-SHO potential,
defined by
\begin{equation}
V(x) =
\left\{
\begin{array}{cc}
m\omega^2 x^2/2 & \mbox{for $x\geq 0$} \\
0               & \mbox{for $x< 0$}
\end{array}
\right.
\label{half_sho}
\end{equation}
and find dramatically different behavior for many sum rules.

We are motivated to study the relation between such parity-related
potentials since the solution space of the parity-restricted version of
a symmetric potential, $V(x) = V(-x)$, consists solely of the
odd energy eigenvalues and eigenstates, with only a trivial change in
normalization. Thus, half of the wavefunctions and
energies will be closely related in the two systems. Since
sum rules involve intricate relationships between energy differences and
matrix elements derived from the wavefunctions,
it is instructive to see how identical sum rules are
realized in two parity-related systems. We find that
many matrix elements can be very different in character between
two such related systems, yielding  sum rule identities which are realized
in novel ways.

For these two systems, we then make contact with semi-classical (WKB-like)
probability distributions (in Sec.~\ref{sec:classical_versus_quantum})
and examine the similarities with the new exact quantum results presented
here in the large $n$ limit.
We also note that a recent examination of the Stark effect in the symmetric
linear potential \cite{robinett_stark}
has found interesting constraints on second-order perturbation theory
sums which are very similar to the energy-weighted sum rules we will
discuss below, and we extend that analysis to the `half'-SHO system
introduced here, allowing for insight into the structure of the Stark effect
in that system.
We begin by briefly reviewing the results of
Ref.~\cite{robinett_stark} in the next section, mostly to establish
notation.

\section{Solutions for the symmetric linear potential}
\label{sec:solutions}

We begin by reviewing the properties of the solutions for the
symmetric linear potential in Eqn.~(\ref{symmetric_linear_potential}),
using a slight modification of the notation in
Ref.~\cite{robinett_stark}. We note that the
only difference in notation with Ref.~\cite{belloni_robinett_airy_zeros}
is in the normalization
of the wavefunctions, corresponding only to a physically irrelevant
difference in phase.

Using the fact that the linear potential admits Airy function solutions,
with different forms for positive and negative values of position, we can
construct piecewise-continuous solutions of the corresponding Schr\"odinger
equation.  Because of the symmetry properties of the potential,
we can also classify the solutions by their parity, with the
odd functions, which must vanish at the origin, related to the solutions
of the quantum bouncer problem. For background on the solutions to
the bouncer problem, see also Refs.~\cite{bouncing_ball},
\cite{vallee} - \cite{goodmanson}.

Specifically, the odd solutions can be written in the form
\begin{equation}
\psi_n^{(-)}(z) =
N_{n}^{(-)}
\left\{
\begin{array}{cc}
Ai(z/\rho - \beta_n) &   \qquad \mbox{for $0 \leq z$} \\
-Ai(-z/\rho - \beta_n) & \qquad \mbox{for $z \leq 0$}
\end{array}
\right.
\label{odd_states}
\end{equation}
where $\rho \equiv (\hbar^2/2mF)^{1/3}$.
The appropriate boundary condition at the origin is that the wavefunction must vanish there which implies that
\begin{equation}
\psi_n^{(-)}(0) = Ai(-\beta_n)
\end{equation}
where we identify $\beta_n = \zeta_n$ where the $-\zeta_n$ are the zeros of $Ai(\zeta)$.
The corresponding combination for even parity solutions is written as
\begin{equation}
\psi_n^{(+)}(x) =
N_{n}^{(+)}
\left\{
\begin{array}{cc}
Ai(z/\rho - \beta_n) & \qquad \mbox{for $0 \leq z$} \\
Ai(-z/\rho - \beta_n) & \qquad \mbox{for $z \leq 0$}
\end{array}
\right.
\label{even_states}
\end{equation}
which at the origin must satisfy $[\psi_n^{(+)}(z\!=\!0)]' = 0$, implying that
$\beta_n = \eta_n$ where the $-\eta_n$ are the zeros of $Ai'(\zeta)$.

Handbook results \cite{stegun}, \cite{book}
give approximations for these zeros, in the large $n$ limit, as
\begin{equation}
\zeta_n \sim \left[\frac{3\pi}{4} (2n-1/2)\right]^{2/3}
\qquad
\quad
\mbox{and}
\quad
\qquad
\eta_n \sim \left[\frac{3\pi}{4} ((2n-1)-1/2)\right]^{2/3}
\label{large_n_expansion}
\end{equation}
where the labeling starts with $n=1$ in both cases.
We note that WKB quantization gives the same results \cite{robinett_stark},
valid for large quantum numbers.

The normalizations required in Eqns.~(\ref{odd_states}) and
(\ref{even_states}) are obtained by using integrals first derived by Gordon \cite{gordon} and
Albright \cite{albright}  collected in the  Appendix in
Sec.~\ref{sec:indefinite}, specifically Eqn.~(\ref{diagonal_0}),
and are given by
\begin{equation}
N_n^{(-)} = \frac{1}{\sqrt{2\rho} \,Ai'(-\zeta_n)}
\qquad
\quad
\mbox{and}
\qquad
\quad
N_n^{(+)} = \frac{1}{\sqrt{2\rho\eta_n} \, Ai(-\eta_n)}
\, .
\label{normalizations}
\end{equation}
We note that the wavefunctions at the origin satisfy
\begin{equation}
\psi_n^{(-)}(0) = 0
\qquad
\quad
\mbox{and}
\quad
\qquad
\psi_n^{(+)}(0) = \frac{1}{\sqrt{2\rho \eta_n}}
\, .
\label{wavefunctions_at_the_origin}
\end{equation}
The energy eigenvalues are then given directly in terms of the
$\zeta_n,\eta_n$ by
\begin{equation}
E_n^{(-)} = {\cal E}_0 \zeta_n
\qquad
\quad
\mbox{and}
\quad
\qquad
E_n^{(+)} = {\cal E}_0 \eta_n
\end{equation}
where $ {\cal E}_0 \equiv \rho F$ and the energy spectrum satisfies
\begin{equation}
E_{1}^{(+)} < E_{1}^{(-)} < E_{2}^{(+)} < E_{2}^{(-)} <  \cdots
\,\,\,  .
\end{equation}

Just as for the bouncer system \cite{goodmanson}, we have
a power-law potential of the form $V_{(k)}(x) = V_0 |x/a|^k$,
and the quantum-mechanical virial theorem,
$\langle\hat{T}\rangle = \frac{k}{2}\langle\hat{V}\rangle$,
is satisfied for $k = 1$ which we can confirm by direct evaluation.
For example, for odd states we have the matrix elements
\begin{eqnarray}
\langle \psi_{n}^{(-)} | V(z) | \psi_{n}^{(-)} \rangle
& = &
\int_{-\infty}^{+\infty}\, |\psi_n^{(-)}(z)|^2\,F|z|\, dz
\nonumber \\
& = &
\frac{2F\rho^2}{2 \rho [Ai'(-\zeta_n)]^2} \int_{0}^{\infty} y\,
[Ai(y-\zeta_n)]^2\,dy
\nonumber \\
& = &
\frac{2}{3} {\cal E}_0 \zeta_n = \frac{2}{3}E_{n}^{(-)} \, ,
\label{virial_theorem}
\end{eqnarray}
using the integral in Eqn.~(\ref{diagonal_1}), with a similar result
for the expectation value of the potential energy for the even states. 
The expectation value of the kinetic energy operator,
\begin{equation}
\langle \psi_n^{(\pm)} | \hat{T} |\psi_n^{(\pm)} \rangle
=
\frac{1}{2m}\langle \psi_n^{(\pm)} | \hat{p}^2 |\psi_n^{(\pm)} \rangle
= \frac{1}{3}E_n^{(\pm)}
\, ,
\label{kinetic_energy_virial}
\end{equation}
can be evaluated by either: (i) using the definition of the Airy differential equation,
$Ai''(\zeta-\beta) = (\zeta-\beta)Ai(\zeta)$, to rewrite the integral in terms of
the one in  Eqn.~(\ref{diagonal_1}) or (ii) using an integration by parts
to move one derivative onto the first wavefunction, and then using the
integral in Eqn.~(\ref{diagonal_derivative}).

The dipole matrix elements needed for many of the best-known sum rules have
a different form than in the quantum bouncer. Using the symmetry of 
the potential, we have
\begin{equation}
\langle \psi_{n}^{(-)} | z | \psi_{n}^{(-)} \rangle
=
\langle \psi_{n}^{(+)} | z | \psi_{n}^{(+)} \rangle
= 0
\, ,
\label{diagonal_dipole_matrix_element}
\end{equation}
while using Eqn.~(\ref{off_diagonal_1}), we find that
\begin{equation}
\langle \psi_{n}^{(-)} | z | \psi_{k}^{(+)} \rangle
= \frac{-2\rho }{\sqrt{\eta_k}\, (\eta_k - \zeta_n)^3}
\,.
\label{off_diagonal_dipole_matrix_element}
\end{equation}
As with all of the matrix element expressions and sum rule identities
we present, we have confirmed these expressions numerically using \mathmath.

In contrast to the quantum bouncer problem \cite{belloni_robinett_airy_zeros} 
where all off-diagonal matrix elements gave terms proportional
to inverse powers of $(\zeta_n - \zeta_k)$, using sum rules
involving only dipole-matrix
elements for the symmetric linear potential
will give constraints on sums of powers of
$1/(\eta_k - \zeta_n)$.  Matrix elements for even powers of $z$,
for example
\begin{equation}
\langle \psi_{n}^{(\pm)} | z^2 | \psi_{n}^{(\pm)} \rangle
\, ,
\end{equation}
will connect even-even and odd-odd states. The corresponding
odd-odd sum rules
will reproduce some of the same identities found in
Ref.~\cite{belloni_robinett_airy_zeros},
while the even-even sum rules will generate new identities,
involving sums over inverse powers of $(\eta_k - \eta_n)$.

\section{New constraints from the symmetric
linear potential}
\label{sec:constraints}

To illustrate the range of new constraints which
the symmetric linear potential places on the $\zeta_n$ and $\eta_n$,
we first consider the most well-known sum rules involving dipole matrix elements.  
The famous Thomas-Reiche-Kuhn (TRK) sum rule \cite{trk_sum_rule} is given by
\begin{equation}
\sum_{k\neq n} (E_k - E_n)|\langle n |z|k\rangle|^2 =
\frac{\hbar^2}{2m}
\label{trk_sum_rule}
\end{equation}
while a completeness relation for the matrix elements of the momentum
operator can be written in terms of dipole matrix matrix elements as
\begin{equation}
\sum_{k\neq n} (E_k - E_n)^2 |\langle n |z|k \rangle|^2
=\frac{\hbar^2}{m^2} \langle n |\hat{p}^2|n\rangle
=\frac{2\hbar^2}{m} \left [ E_n - \langle n|V(z)|n\rangle \right ]
\label{second_power_momentum_sum_rule}
\,.
\end{equation}
We note that since the states $n$ and $k$ that contribute
to the sum rule will be of opposite parity, the
$k\neq n$ restriction is unnecessary and we can sum over all
intermediate state $k$ labels.

Two other sum rules involving dipole matrix elements have been
discussed by Bethe and Jackiw \cite{bethe_intermediate},
\cite{jackiw_sum_rules}, namely
\begin{equation}
\sum_{k}(E_k - E_n)^3 |\langle n |z| k \rangle|^2
= \frac{\hbar^4}{2m^2} \left\langle n \left|
\frac{d^2 V(z)}{dz^2} \right| n \right\rangle
\label{first_potential_sum_rule}
\end{equation}
and
\begin{equation}
\sum_{k} (E_k - E_n)^4 |\langle n |z| k \rangle|^2
= \frac{\hbar^4}{m^2} \left\langle n \left|\left(\frac{dV(z)}{dz}\right)^2
\right| n \right\rangle
\label{second_potential_sum_rule}
\end{equation}
which are sometimes called the {\it force times momentum} and
{\it force-squared} sum rules, respectively.
We note that not all such sum rules necessarily lead
to convergent expressions. In Ref.~\cite{belloni_robinett_airy_zeros}, 
Eqns.~(\ref{first_potential_sum_rule}) and (\ref{second_potential_sum_rule})
could not be applied in the quantum bouncer system due to the
discontinuous nature of the potential in Eqn.~(\ref{bouncer_potential}).
In the case of the symmetric linear potential, the system is defined by a
continuous potential and even the discontinuous derivative will lead to a convergent result in the
corresponding sum rule in Eqn.~(\ref{first_potential_sum_rule}).
These results regarding convergence are consistent with the
form of the dipole matrix elements in
Eqn.~(\ref{off_diagonal_dipole_matrix_element}),  where their values
for fixed $k$ or $n$ clearly decrease more quickly with
increasing $n$ or $k$ than their counterparts in the bouncer system,
indicating better convergence.

When we fix the state labeled $n$ as being one of the odd solutions and
sum over $k$ values corresponding to the even states,  the
$1/\sqrt{\eta_k}$ factors remain inside the summation, giving sums of the form
\begin{equation}
T_{p}(n) \equiv \sum_{\textrm{all}\, k} \frac{1}{\eta_k (\eta_k - \zeta_n)^p}
\,.
\label{T_definition}
\end{equation}
In contrast, when the even states are fixed, switching the roles of $n$ and
$k$ in Eqn.~(\ref{off_diagonal_dipole_matrix_element}), then the
common $1/\sqrt{\eta_n}$ factor can be removed from inside
the summation and eventually moved to the right-hand side
of the identity involved. In those cases, we find expressions of the form
\begin{equation}
U_{p}(n) \equiv \sum_{\textrm{all}\, k} \frac{1}{\eta_n(\zeta_k - \eta_n)^p}
= \left(\frac{1}{\eta_n}\right)
\sum_{\textrm{all}\, k} \frac{1}{(\zeta_k - \eta_n)^p}
\equiv \frac{1}{\eta_n} \tilde{U}_p(n)
\label{new_U}
\end{equation}
where
\begin{equation}
\tilde{U}_p(n)
\equiv
\sum_{\textrm{all}\, k} \frac{1}{(\zeta_k - \eta_n)^p}
\,.
\end{equation}
The related quantity where we sum over $k$ values corresponding to
odd states, namely
\begin{equation}
\tilde{T}_p(n)
\equiv
\sum_{\textrm{all}\, k} \frac{1}{(\eta_k - \zeta_n)^p}
\,.
\end{equation}
can be obtained by using the definition in Eqn.~(\ref{T_definition}) to
write
\begin{equation}
\tilde{T}_p(n) = \zeta_n\, T_p(n) + T_{p-1}(n)
\label{new_T}
\,.
\end{equation}
For completeness, we recall from Ref.~\cite{belloni_robinett_airy_zeros}
that for the quantum bouncer system constraints on sums were of the form
\begin{equation}
S_{p}(n) \equiv \sum_{k\neq n} \frac{1}{(\zeta_k-\zeta_n)^{p}}
\, .
\end{equation}
We will focus on evaluation of the $T_p(n)$ and $U_p(n)$, noting that
it is straightforward to obtain the related $\tilde{T}_p(n),\tilde{U}_p(n)$
for comparison to $S_p(n)$, using
Eqns.~(\ref{new_U}) and (\ref{new_T}).

While not formally a sum rule,
we note that the completeness relation for dipole matrix elements
can be written as
\begin{equation}
\sum_{k\neq n} |\langle n |z|k \rangle|^2
+ \langle n|z|n\rangle =
\sum_{\textrm{all}\, k} |\langle n |z|k \rangle|^2 = \langle n |z^2|n \rangle
\label{x_completeness}
\end{equation}
and that because of the parity of the potential the `diagonal' term is in fact zero. While not a
sum rule in the classic sense, it will also provide novel constraints on
the $Ai$ and $Ai'$ zeros.

For the sum rules in Eqns.~(\ref{trk_sum_rule}) -
(\ref{second_potential_sum_rule}) and (\ref{x_completeness}),
we can now fix a state of definite parity,
either $\psi_n^{(+)}(z)$ or $\psi_n^{(-)}(z)$, and  then sum over
the states of opposite parity, namely the $\psi_k^{(-)}(z)$ or
$\psi_k^{(+)}(z)$, respectively. In this way, we obtain two different
constraints for each quantum-mechanical sum rule.

For example, starting with the TRK sum rule in Eqn.~(\ref{trk_sum_rule})
we find the relations
\begin{equation}
T_5(n) \equiv  \sum_{\textrm{all} \, k} \frac{1}{\eta_k(\eta_k - \zeta_n)^5}  =  \frac{1}{4}
= \left(\frac{1}{\eta_n}\right) \sum_{\textrm{all} \, k} \frac{1}{(\zeta_k - \eta_n)^5}
\equiv U_5(n)
\label{trk_result}
\end{equation}
by using odd ($\psi_n^{(-)})$ and even ($\psi_n^{(+)}$) states respectively.
For comparison, we note that the constraint arising in this case for the
quantum bouncer is
\begin{equation}
S_3(n) = \sum_{k\neq n} \frac{1}{(\zeta_k-\zeta_n)^{3}} = \frac{1}{4}
\, .
\end{equation}

From the momentum-completeness sum rule in
Eqn.~(\ref{second_power_momentum_sum_rule}), we use the evaluation of
the the $\hat{p}^2$ operator from Eqn.~(\ref{kinetic_energy_virial})
and find
\begin{eqnarray}
T_{4}(n) & = & \sum_{\textrm{all}\, k} \frac{1}{\eta_k(\eta_k - \zeta_n)^4}
= \frac{\zeta_n}{3}
\label{momentum_completeness_1} \\
U_{4}(n) & = & \left(\frac{1}{\eta_n}\right)\sum_{\textrm{all} \, k} \frac{1}{(\zeta_k - \eta_n)^4}
= \frac{\eta_n}{3}
\label{momentum_completeness_2}
\end{eqnarray}
while the corresponding sum rule constraint for the bouncer is
\begin{equation}
S_3(n) = \sum_{k\neq n} \frac{1}{(\zeta_k-\zeta_n)^{2}} = \frac{\zeta_n}{3}
\, .
\end{equation}

For the sum rule in Eqn.~(\ref{second_potential_sum_rule}), 
the fact that $[V'(z)]^2 = (\pm F)^2 = F^2$ for both positive and
negative values of $z$, yields
\begin{equation}
T_{2}(n)  =  \sum_{\textrm{all} \, k} \frac{1}{\eta_k(\eta_k - \zeta_n)^2}
= 1
=
\left( \frac{1}{\eta_n}\right)\sum_{\textrm{all} \, k} \frac{1}{(\zeta_k - \eta_n)^2}
= U_{2}(n)
\,.
\label{force_squared_result}
\end{equation}
The convergence of these  sums is clear since
the large $k$ behavior of the $\zeta_k \sim k^{2/3}$ implies that each
term scales as $k^{-4/3}$.
Thus, $p=2$ is the smallest power which will
lead to a convergent sum rule. For the case of the
quantum bouncer, this sum rule did not  converge.

For the sum rule in Eqn.~(\ref{first_potential_sum_rule}) we require the
value of $V''(z)$. Given the cusp in the definition of $V(z)$, we find that
$V''(z) = 2F\delta(z)$ so that the right-hand side of
Eqn.~(\ref{first_potential_sum_rule}) becomes
\begin{equation}
\frac{\hbar^4F}{m^2} |\psi_n^{(\pm)}(0)|^2
\, .
\label{force_times_momentum_result}
\end{equation}
We can then use the results of Eqn.~(\ref{wavefunctions_at_the_origin})
and immediately see that
\begin{eqnarray}
T_{3}(n) & = & \sum_{\textrm{all} \, k} \frac{1}{\eta_k(\eta_k - \zeta_n)^3}
= 0 \\
U_{3}(n) & = & \left(\frac{1}{\eta_n}\right)
\sum_{\textrm{all} \, k} \frac{1}{(\zeta_k - \eta_n)^3}  = \frac{1}{2}
\,  ,
\end{eqnarray}
and note that the form of the right-hand side of this sum-rule constraint varies in
form between the odd and even states. The only other case of which we are aware where the {\it forces times
momentum} sum rule gives a similar result,
namely where the sum rule depends on the wavefunction at the origin,
is for the Coulomb problem \cite{bethe_intermediate}. For that system, we have
$\nabla^2 V({\bf r}) = -e \delta({\bf r})$ and the sum rule gives a non-zero
result for $S$-wave ($l=0$ states) only.

To evaluate the constraint arising from the
$z$-completeness relation in Eqn.~(\ref{x_completeness}), we require the
expectation values of $z^2$ in the even and odd states. These can be
derived by using the integral in Eqn.~(\ref{diagonal_2}) to obtain
\begin{equation}
\langle \psi_{n}^{(-)} | z^2 | \psi_{n}^{(-)} \rangle
 =
\rho^2 \left(\frac{8 \zeta_n^2}{15} \right)
\qquad
\mbox{and}
\qquad
\langle \psi_{n}^{(+)} | z^2 | \psi_{n}^{(+)} \rangle
 = \rho^2 \left( \frac{8 \eta_n^2}{15} + \frac{1}{5\eta_n} \right)
\label{second_moments}
\end{equation}
where we once again notice the difference in form between the results
for the two parities. Using these results, we find the following sum rules
\begin{eqnarray}
T_6(n) = \sum_{\textrm{all} \, k} \frac{1}{\eta_k(\eta_k - \zeta_n)^6}
& = & \frac{2\zeta_n^2}{15} \\
U_6(n) =  \left(\frac{1}{\eta_n}\right)
\sum_{\textrm{all} \, k} \frac{1}{(\zeta_k - \eta_n)^6}
& = & \frac{2\eta_n^2}{15} + \frac{1}{20\eta_n}
\,.
\end{eqnarray}
The corresponding result for the quantum bouncer
\cite{belloni_robinett_airy_zeros} has a non-zero
`diagonal' term, as in Eqn.~(\ref{x_completeness}),
which leads to the result
\begin{equation}
S_4 = \sum_{k\neq n} \frac{1}{(\zeta_k-\zeta_n)^{4}} = \frac{\zeta_n^2}{45}
\end{equation}

It has been stressed \cite{belloni_robinett_sum_rules} that the form of the
standard second-order perturbation theory result, namely
\begin{equation}
E_n^{(2)} =
\sum_{k \neq n}
\frac{|\langle n |\overline{V}(x)|k\rangle|^2}
{(E_n^{(0)} - E_k^{(0)})}
\label{general_second_order_shift}
\end{equation}
can also be thought of as an energy-weighted sum rule. 
Jackiw \cite{jackiw_sum_rules} discussed energy-difference weighted sum rules, 
containing factors such as $(E_k-E_n)^p$
with negative values of $p$. This concept was
applied in Ref.~\cite{belloni_robinett_airy_zeros}
to obtain another independent constraint
on the $S_p(n)$ for the quantum bouncer by considering the perturbing
effect  of an additional constant force (linear field) via the Stark effect.

The Stark effect for the symmetric linear potential has recently
been discussed \cite{robinett_stark} where a straightforward expansion of the
exact eigenvalue condition was shown to led to closed-form expressions for
the second-order energy shifts. If the perturbing potential is defined as
$\overline{V}(z) = \overline{F}z$, then the results for the Stark shifts for
the odd- and even-parity states respectively are found to be
\begin{equation}
E_n^{(-,2)}  =  - \frac{7}{9} \left(\frac{\overline{F}}{F}\right)^2\,
E_n^{(-)}
\qquad
\quad
\mbox{and}
\qquad
\quad
E_n^{(+,2)}  =  - \frac{5}{9} \left(\frac{\overline{F}}{F}\right)^2\,
E_n^{(+)}
\, .
\label{symmetric_linear_stark_shift}
\end{equation}
Using the expression in Eqn.~(\ref{general_second_order_shift}),
we find the the second-order shifts given by perturbation theory are
\begin{eqnarray}
E_n^{(-,2)} & = &
- 4\left(\frac{\overline{F}}{F}\right)^2 {\cal E}_0
\left[\sum_{\textrm{all} \, k} \frac{1}{\eta_k(\eta_k - \zeta_n)^7}\right]
\\
E_n^{(+,2)} & = &
- 4\left(\frac{\overline{F}}{F}\right)^2 {\cal E}_0
\left[\sum_{\textrm{all} \, k} \frac{1}{\eta_n(\zeta_k - \eta_n)^7}\right]
\end{eqnarray}
which gives two new constraints,
\begin{eqnarray}
T_7(n) & = & \sum_{\textrm{all} \, k} \frac{1}{\eta_k(\eta_k - \zeta_n)^7} =
+\frac{7 \zeta_n}{36} \\
U_7(n) & = &
\left(\frac{1}{\eta_n}\right)\sum_{\textrm{all} \, k} \frac{1}{(\zeta_k - \eta_n)^7} =
+\frac{5 \eta_n}{36}
\end{eqnarray}
which can be compared to the the quantum bouncer,
\begin{equation}
S_5(n) = \sum_{k \neq n}\frac{1}{(\zeta_k - \zeta_n)^5} = \frac{\zeta_n}{36}
\, .
\end{equation}

Our earlier discussion of sum rules for the quantum bouncer
\cite{belloni_robinett_airy_zeros} allowed for
a systematically calculable hierarchy of constraints on sums of inverse powers
of $(\zeta_k-\zeta_n)$. This was made possible by repeated use of the
commutation relation $[x^q,\hat{p}] = i q\hbar
x^{q-1}$ to recursively obtain sums over matrix elements for higher and
higher moments. For the symmetric potential, where matrix elements of
odd powers of $z$ vanish, that connection is lost and we know of no 
strategy to generate all of the $T_p(n)$ and $U_p(n)$ in a systematic
way. We can, however, make use of the two relations
\begin{eqnarray}
\sum_{all \, k} \langle n | x^q| k\rangle \langle k |x| n\rangle
& = & \langle n |x^{q+1}| n \rangle
\label{odd_extend_1}\\
\sum_{all \, k} (E_k - E_n)\langle n | x^q| k\rangle \langle k |x| n\rangle
& = & q \left(\frac{\hbar^2}{2m}\right)\langle n |x^{q-1}| n \rangle
\label{odd_extend_2}
\end{eqnarray}
for $q$ odd to generate an infinite number of new constraints on inverse powers
of $(\eta_n - \zeta_k)$ similar to the ones derived already, requiring
only the expectation values on the right-hand sides, which can in turn be
evaluated using the recursion relations in Appendix~\ref{sec:recursion}.

The sum rules involving dipole matrix elements for the
symmetric linear potential, using Eqns.~(\ref{diagonal_dipole_matrix_element}) and
(\ref{off_diagonal_dipole_matrix_element}), give
constraints on sums of inverse powers of $(\eta_k - \zeta_n)$. In order
to obtain constraints on differences of just the zeros of
$Ai'(\zeta)$ separately, namely the $(\eta_k - \eta_n)$,
we need to consider matrix elements which are non-vanishing
for even-even states. For example, because of the parity constraints in
the symmetric linear potential, the so-called monopole sum rule
\cite{bohigas}
\begin{equation}
\sum_{k \neq n} (E_k - E_n)|\langle n |z^2|k\rangle|^2 =
\frac{2\hbar^2}{m} \langle n |z^2|n \rangle
\label{monopole_sum_rule}
\end{equation}
will connect only odd-odd and even-even states; in this case the
constraint that $k \neq n$ is indeed required. For the odd-states,
the corresponding constraint on the $\zeta_n$ was found in
Ref.~\cite{belloni_robinett_airy_zeros} to be
\begin{equation}
S_7(n) = \sum_{k \neq n} \frac{1}{(\zeta_k - \zeta_n)^7} =
\frac{\zeta_n^2}{270}
\,.
\end{equation}
For the even-even case, we can use the new matrix element result
in Eqn.~(\ref{off_diagonal_2}) to evaluate
\begin{equation}
\langle \psi_n^{(+)} | z^2 | \psi_k^{(+)} \rangle
= - \frac{12 (\eta_n + \eta_k)}{\sqrt{\eta_n \eta_k} (\eta_n - \eta_k)^4}
\end{equation}
which gives a more complicated constraint, namely
\begin{equation}
\sum_{k \neq n} \frac{(\eta_n+\eta_k)^2}{\eta_n \eta_k (\eta_n -\eta_k)^7}
= \frac{1}{36}\left( \frac{8 \eta_n^2}{15} + \frac{1}{5\eta_n}
\right)
\,.
\end{equation}
Once again, we have checked all of these results numerically using
\mathmath.

The new classes of constraints on the zeros of $Ai$ and $Ai'$ derived
here, from application of sum rules for the symmetric linear potential,
are seen to be qualitatively similar to those obtained from the
parity-restricted version of this potential, namely the quantum bouncer.
In contrast, the nature of the mathematical relations dictated
by sum rules applied to the parity-restricted version of the harmonic
oscillator are qualitatively very different, as we will see in
Sec.~\ref{sec:half_sho}, after first reviewing sum rules in the
standard oscillator system.

\section{Review of sum rules for the harmonic oscillator}
\label{sec:sho}

Before examining the parity-restricted version of the harmonic oscillator,
we briefly review the solutions for the the familiar oscillator
potential, as well as
the structure of the quantum mechanical sum rules for this system.
The solutions for the Schr\"{o}dinger equation
\begin{equation}
- \frac{\hbar^2}{2m} \frac{d^2 \psi_n(x)}{dx^2}
+ \frac{1}{2} m\omega^2 x^2\,\psi_n(x) = E_n\psi_n(x)
\end{equation}
can be written in the form
\begin{equation}
\psi_n(x) = \frac{c_n}{\sqrt{\beta}}\, H_n(y) \, e^{-y^2/2}
\qquad
\quad
\mbox{where}
\quad
\qquad
c_n = \frac{1}{\sqrt{2^n n! \sqrt{\pi}}}
\label{sho_states}
\end{equation}
where
\begin{equation}
x = \beta y
\qquad
\quad
\mbox{and}
\qquad
\quad
\beta \equiv \sqrt{\frac{\hbar}{m\omega}}
\end{equation}
with $y$ dimensionless and the $H_n(y)$ are the Hermite polynomials.
 The corresponding energy eigenvalues  are
\begin{equation}
E_n = (n+1/2)\hbar \omega
\qquad
\quad
\mbox{or}
\qquad
\quad
\epsilon_n \equiv \frac{2E_n}{\hbar \omega} = 2n+1
\end{equation}
with $n=0,1,...$
so that the differential equation in dimensionless form can be written as
\begin{equation}
\psi_n''(y) = (y^2 - \epsilon_n) \psi_n
\, .
\label{dimensionless_sho}
\end{equation}
The solutions have parity given by $P_n = (-1)^n$ and the expectation
values of any odd power of $x$ vanishes,
so that, for example, $\langle n |x^{2p+1}|n\rangle = 0$.

Both $x$ and $\hat{p}$ can be written in terms of raising and lowering
operators, $\hat{A}$ and $\hat{A}^{\dagger}$,
and hence the multipole matrix elements exhibit an exceptionally simple
structure leading to absolute  selection rules.
For example, we have from standard textbooks the relations
\begin{eqnarray}
\langle n |x| k \rangle
& = & \frac{\beta}{\sqrt{2}}
\left\{ \delta_{n,k-1} \sqrt{k} + \delta_{n,k+1} \sqrt{k+1}\right\}
\label{full_sho_1} \\
\langle n |x^2 | k \rangle
& = & \frac{\beta^2}{2}
\left\{
\delta_{n,k-2} \sqrt{k(k-1)}
+ \delta_{n,k} (2k+1)
+ \sqrt{n,k+2} \sqrt{(k+1)(k+2)} \right\}
\label{full_sho_2} \\
\langle n |x^3 | k \rangle
& = & \frac{\beta^3}{2\sqrt{2}}
\left\{
\delta_{n,k-3} \sqrt{k(k-1)(k-2)}
+ 3\delta_{n,k-1} k^{3/2}
+ \delta_{n,k+1} (k+1)^{3/2} \right. \nonumber \\
& &
\qquad \qquad \qquad
\left.
+ \delta_{n,k+3} \sqrt{(k+1)(k+2)(k+3)}
\right\}
\, .
\label{full_sho_3}
\end{eqnarray}
These, and similar expressions for matrix elements of powers $\hat{p}$,
imply that all of the familiar sum rules discussed above
will be `super-convergent', namely
that the relevant infinite sums will actually be saturated by a finite number
of terms, and hence satisfied in a trivial way. We will find
that the situation is dramatically different in the case of the oscillator
restricted to the half-line, at least for the matrix elements of odd
values of $x$, and that is the subject of the next section. The matrix
element relations for even powers of $x$, including
Eqn.~(\ref{full_sho_2}) will still be relevant for the restricted oscillator case with minor relabeling.

\section{The parity-restricted harmonic oscillator}
\label{sec:half_sho}

The solutions to the quantum mechanical problem of a particle
in the potential in Eqn.~(\ref{half_sho}) are easily obtained
from the odd-parity solutions in Eqn.~(\ref{sho_states}).
In dimensionless notation, we have
\begin{equation}
\tilde{\psi}_n(y) =
\left\{
\begin{array}{cc}
\sqrt{2} \, \psi_{2n+1}(y) & \mbox{for $y \geq  0$} \\
0  & \mbox{for $y\leq 0$}
\end{array}
\right.
\label{half_sho_solutions}
\end{equation}
where $n=0,1,2,...$ for the `half'-SHO states $\tilde{\psi}_n(y)$ associated
with the odd solutions of the oscillator, and an appropriate change in
overall normalization.  We will henceforth use integrals over either
the dimensional $x$ or dimensionless $y$ variable as deemed most useful.
We note for future reference that
the derivatives of the solutions at the origin necessary for matrix
element calculations are given by
\begin{equation}
\tilde{\psi}_n'(0) = \sqrt{2} \, c_{2n+1} \, H'_{2n+1}(0)
=
\frac{(-1)^n \, 2}{\sqrt{(2n+1)! \sqrt{\pi}}} (2n+1)!!
=
\frac{(-1)^n 2}{2^n n!} \sqrt{\frac{(2n+1)!}{\sqrt{\pi}}}
\end{equation}
and
\begin{equation}
[\tilde{\psi}_n'(0)]^2 = \frac{4}{2^{2n} (n!)^2} \frac{(2n+1)!!}{\sqrt{\pi}}
= \frac{1}{\sqrt{\pi}}
\left[\frac{4(2n+1)!}{2^{2n} (n!)^2}\right]
= \frac{D_n}{\sqrt{\pi}}
\end{equation}
where
\begin{equation}
D_n \equiv
\frac{4(2n+1)!}{2^{2n} (n!)^2}
\label{definition_of_d_n}
\,.
\end{equation}
Using the Stirling approximation, $n! \sim \sqrt{2\pi n} (n/e)^n$, we
have for large $n$
\begin{equation}
D_n \rightarrow 8 \sqrt{\frac{n}{\pi}}
\end{equation}
which will be useful in examining the semi-classical limit.

The quantized energies for the `half'-SHO
are then given by $\tilde{E}_n = E_{2n+1} = (2n+3/2)\hbar
\omega$ or $\tilde{\epsilon}_n = 4n+3$.  The $\tilde{\psi}_n(y)$
form an orthogonal set which can be seen explicitly by using the recursion relation
derived for oscillator solutions in Appendix~\ref{sec:recursion}.
Using $f(y) = 1$ in
Eqn.~(\ref{full_sho_recursion_relation}), we find that
\begin{equation}
(\tilde{\epsilon}_n - \tilde{\epsilon}_m)^2
\int_{0}^{\infty} \tilde{\psi}_n(y) \, \tilde{\psi}_m(y) \, dy = 0
\end{equation}
so that $\langle \tilde{\psi}_m | \tilde{\psi}_n \rangle = 0$ if $n\neq m$.

For the important dipole matrix elements, we use $f(y) = y$, so that
$f'(0) = 1$ and find that
\begin{equation}
\langle \tilde{\psi}_m |y| \tilde{\psi}_n \rangle =
\int_{0}^{\infty} y\, \tilde{\psi}_n(y) \, \tilde{\psi}_m(y) \, dy
= - \frac{\tilde{\psi}_n'(0) \tilde{\psi}_m'(0)}{2[4(n-m)^2-1]}
\label{half_sho_dipole_matrix_element}
\end{equation}
since $\tilde{\epsilon}_{n} = 4n+3$.

For the special case of $n=m$, the expectation values in the state
$\tilde{\psi}_n$ are therefore given by
\begin{equation}
\langle \tilde{\psi}_n |x| \tilde{\psi}_n \rangle = \frac{D_n}{2\sqrt{\pi}}
\beta
\quad
\longrightarrow
\quad
\frac{4\sqrt{n}}{\pi}\beta
\label{large_n_quantum}
\end{equation}
in the large $n$ limit. We can confirm this using the classical (WKB-like)
probability distribution for the `half'-SHO, namely
\begin{equation}
P_{CL}^{(n)}(x) = \frac{2}{\pi} \frac{1}{\sqrt{A^2-x^2}}
\qquad
\mbox{where}
\qquad 0 \leq x \leq A_n
\end{equation}
where the upper classical turning point, $A_n$, is given by
\begin{equation}
  \frac{1}{2} m \omega^2 A_n^2 = E_n = \hbar\omega (2n+3/2)
\quad
\qquad
\mbox{or}
\qquad
\quad
A_n = \beta \sqrt{4n+3}
\,.
\end{equation}
The classical expectation value is then
\begin{equation}
\langle n |x| n \rangle_{CL}
\equiv
\int_{0}^{A_n} x\, P_{CL}^{(n)}(x)\,dx
=\frac{2A_n}{\pi} = \frac{2\sqrt{4n+3}\,\beta}{\pi}
\quad
\longrightarrow
\quad
\left(\frac{4\sqrt{n}}{\pi}\right)\beta
\label{classical_half_sho_result}
\end{equation}
for large $n$, which agrees with Eqn.~(\ref{large_n_quantum}).

The dipole matrix elements in Eqn.~(\ref{half_sho_dipole_matrix_element})
are clearly very different from the `full'-SHO
case, with each state being connected to all of the
others in a very simple form,
proportional to the derivative of the wavefunction at the origin for each
state, and with an `energy denominator' factor. This is the only
example of a dipole matrix element in a model 1D system of which we are
aware for which the diagonal ($n=m$) and off-diagonal ($n\neq m$) cases
can be written with the same simple expression. The corresponding sum rules
will then be realized in a completely different manner than the
super-convergent form for the more familiar oscillator.

For example, the TRK sum rule in Eqn.~(\ref{trk_sum_rule}) gives the constraint
\begin{equation}
D_n \sum_{k \neq n} \frac{(k-n) D_k}{[4(n-k)^2-1]^2}
= \pi
\label{half_sho_1}
\end{equation}
where the $D_n$ are given in Eqn.~(\ref{definition_of_d_n}). In a similar
way, the $x$-completeness relation of Eqn.~(\ref{x_completeness}) gives
\begin{equation}
D_n \sum_{k \neq n} \frac{D_k}{[4(n-k)^2-1]^2} = (8n+6)\pi
\end{equation}
where we use the fact that
\begin{equation}
\langle \tilde{\psi}_n |y^2 | \tilde{\psi}_n \rangle = (2n+3/2)
\label{half_sho_2}
\end{equation}
making use of the matrix element in Eqn.~(\ref{full_sho_2}) with a suitable
relabeling. One can then combine Eqns.~(\ref{half_sho_1}) and
(\ref{half_sho_2}) to obtain the constraint
\begin{equation}
D_n \sum_{k \neq n} \frac{k\,D_k}{[4(n-k)^2-1]^2}
= (4n+1)(2n+1)\pi
\, .
\label{half_sho_3}
\end{equation}
Looking at the $n$ and $k$ dependence of the dipole matrix elements, one can confirm that the sum rules in
Eqns.~(\ref{first_potential_sum_rule})
and (\ref{second_potential_sum_rule}) which depend on derivatives of
$V(x)$ are not convergent, this result is consistent with the nature of the
discontinuous potential.

Matrix elements involving even powers of $x$ can be written in terms
 of the results for the `full'-SHO and have the same simple `nearby
neighbor' selection rule structure. For example, a simple relabeling
of Eqn.~(\ref{full_sho_2}) gives
\begin{equation}
\langle \tilde{\psi}_n |y^2| \tilde{\psi}_k \rangle
=
\frac{1}{2}
\left\{
\delta_{n,k-1} \sqrt{(2k+1)(2k)}
+ \delta_{n,k}(4k+3)
+ \delta_{n,k+1} \sqrt{(2k+2)(2k+3)}
\right\}
\end{equation}
and the monopole sum rule in Eqn.~(\ref{monopole_sum_rule}) is saturated
by a finite number of terms.
Matrix element relations using odd powers of $x$, such as those in
Eqns.~(\ref{odd_extend_1}) and (\ref{odd_extend_2}) give increasing
complicated relations involving the $D_{k,n}$ since the matrix elements
of $x^{2q+1}$ all are proportional to $\langle \tilde{\psi}_n |x|
\tilde{\psi}_k \rangle$. For example, using the recursion relation
in Eqn.~(\ref{full_sho_recursion_relation}), we find that
\begin{eqnarray}
\langle \tilde{\psi}_n |y^3| \tilde{\psi}_k \rangle
& = &
\frac{-6 (2n+2k+3)}{[4(n-k)^2 - 9]}\,
\langle \tilde{\psi}_n |y| \tilde{\psi}_k \rangle \nonumber \\
& = &
\frac{-3(2n+2k+3)}{[4(n-k)^2-9][4(n-k)^2-1]}
\left(\frac{\tilde{\psi}_n'(0) \tilde{\psi}_k'(0)}{2}\right)
\end{eqnarray}
One can continue to generate increasingly complex constraints by
use of Eqns.~(\ref{odd_extend_1}) and (\ref{odd_extend_2}) for the non-trivial
case of matrix elements of odd powers of $x$.

The realization of the sum-rules for the
parity-restricted version of the oscillator are completely
different than the trivial manner in which they are satisfied for the
ordinary oscillator system and generate an infinite number of constraints
on the $D_n$.

\section{Classical versus quantum mechanical results}
\label{sec:classical_versus_quantum}

We have seen that some of the matrix element expressions and/or sum
rules for the symmetric linear potential
yield identical results when considering both the even and odd cases,
with only the substitution $\zeta_n \leftrightarrow \eta_n$ required: 
the virial theorem result
in Eqn.~(\ref{virial_theorem}) and the TRK, momentum completeness,
and {\it force-squared} sum rules in Eqns.~(\ref{trk_result}),
(\ref{momentum_completeness_1}),
(\ref{momentum_completeness_2}), and (\ref{force_squared_result}).
For others, however, such as the second-moment expectation values in
Eqn.~(\ref{second_moments}) or the {\it force times momentum} result 
one finds slightly different results.

To examine these small differences, we will consider the
classical underpinnings of some of the results which arise in the
evaluation of expectation values. By generalizing the
recursion relation derived by Goodmanson \cite{goodmanson}
in Sec.~\ref{sec:recursion} one can generate the expectation values and
off-diagonal matrix elements of any power of $z$.

For example, from Ref.~\cite{belloni_robinett_airy_zeros}
we know that expectation values of $z^p$ for the quantum bouncer
solutions are given by
\begin{eqnarray}
\langle n | y|n \rangle & = &\frac{2\zeta_n}{3}
\label{old_diagonal_1} \\
\langle n | y^2 |n \rangle & = & \frac{8 \zeta_n^2}{15}
\label{old_diagonal_2} \\
\langle n | y^3 |n \rangle & = & \frac{16\zeta_n^3}{35}
+ \frac{3}{7}
\label{old_diagonal_3} \\
\langle n | y^4|n \rangle & = &
\frac{128\zeta_n^4}{315} + \frac{80\zeta_n}{63}
\label{old_diagonal_4} \\
\langle n | y^5|n \rangle & = &
\frac{256 \zeta_n^5}{693} + \frac{1808\zeta_n^2}{3003}
\label{old_diagonal_5}
\end{eqnarray}
where $y = z/\rho$ and we only show the dimensionless results. 
Each result has a highest-order term of order $(\zeta_n)^p$,
followed by sub-leading terms of order $\zeta^{p-3k}_n$, if present at all.
Using the  large-$n$ expansion of the $\zeta_n$ in
Eqn.~(\ref{large_n_expansion}), we see that the sub-leading terms
are a factor of $(\zeta_n)^{-3k} \sim n^{-k}$ smaller and so become
negligible in the classical limit. This suggests that the leading terms
are indeed what one would expect from a purely classical probability
density.

To confirm this, we note that for the quantum bouncer we have
\begin{equation}
P_{CL}^{(n)}(z) = \frac{1}{2\sqrt{A_n(A_n-z)}}
\label{classical_bouncer_distribution}
\end{equation}
where $A_n$ is the upper classical turning point, defined by $E_n = FA_n$.
If we equate the total energy with the quantum mechanical result
$E_n = (F\rho)\zeta_n$ and write $A_n = \rho \zeta_n$,
the classical probability density reduces to
\begin{equation}
P_{CL}^{(n)}(z) = \frac{1}{2\rho \sqrt{\zeta_n(\zeta_n-z/\rho)}}
\label{purely_classical_bouncer}
\,.
\end{equation}
We briefly discuss, in Sec.~\ref{sec:semiclassical_limit},
how this classical distribution can also be extracted
directly from the large $n$ limit of the exact quantum solutions.

The expectation values of moments of position are then given by
\begin{eqnarray}
\langle n |z^p | n \rangle_{CL}
& = &\int_{0}^{A_n} z^p \, P_{CL}^{(n)}(z)\,dz \nonumber \\
& = & \rho^p \int_{0}^{\zeta_n} \frac{z^p}{2\sqrt{\zeta_n (\zeta_n -z)}}
\, dz \nonumber \\
& = & \frac{\rho^p \zeta_n^p}{2} \int_{0}^{1} \frac{y^p}{\sqrt{1-y}}\,dy
\nonumber \\
& = & \frac{(\rho \zeta_n)^p}{2}
B(p+1,1/2) \nonumber \\
& = &
\frac{(\rho \zeta_n)^p}{2}
\frac{\Gamma(1+p)\Gamma(1/2)}{\Gamma(p+3/2)}
\label{classical_result_for_leading_term}
\end{eqnarray}
and this expression agrees with the leading order terms in
Eqns.~(\ref{old_diagonal_1}) - (\ref{old_diagonal_5}) up through $p=5$.
Using the
recursion relation of Goodmanson, reviewed and extended in the
Appendix, we can assume generally that the behavior of the leading
term for the expectation values can be written as $\langle n |x^p |n
\rangle = A_n^p$ and the recursion relation requires that
\begin{equation}
A_n^q = \frac{2q \zeta_n}{(2q+1)} A_n^{q-1}
\qquad
\quad
\mbox{or}
\qquad
\quad
A_n^{p} = \frac{2^p p!}{(2p+1)!!} \zeta_n^p
\end{equation}
which agrees with the classical result in
Eqn.~(\ref{classical_result_for_leading_term}) for all $p$ values.

For the symmetric linear potential, the expectation values
for odd powers of $x$ vanish, but the integrals in
Eqns.~(\ref{old_diagonal_1}) - (\ref{old_diagonal_5}) are still useful.
For example, for the expectation value of $V(z) = F|z|$, we require the result
in Eqn.~(\ref{old_diagonal_1}). Moreover, because of the piecewise
continuous definition of the wavefunctions in Eqns.~(\ref{odd_states})
and (\ref{even_states}), the integrals obtained by using the
recursion relations in Appendix~\ref{sec:recursion} are
necessarily defined over the interval $(0,\infty)$ and then extended
over all space, so intermediate results for integrals over the half-line
which eventually  vanish due to parity constraints can still be useful.

To compare the leading and sub-leading contributions to integrals used for
the expectation values for odd and even states, we can compare the results
in Eqns.~(\ref{old_diagonal_1}) - (\ref{old_diagonal_5}) to similar results
for the even states. These can be defined by using the states in
Eqn.~(\ref{even_states}), integrated over positive values of $x$ and
normalized as for the quantum bouncer. For those cases we find
\begin{eqnarray}
\langle n |y| n \rangle & = & \frac{2\eta_n}{3} \\
\langle n |y^2| n \rangle & = & \frac{8\eta_n^2}{15} + \frac{1}{4\eta_n} \\
\langle n |y^3| n \rangle & = &  \frac{16\eta_n^3}{35} + \frac{3}{5} \\
\langle n |y^4| n \rangle & = &  \frac{128 \eta_n^4}{315} +
\frac{64 \eta_n}{45} \\
\langle n |y^5| n \rangle & = & \frac{256 \eta_n^5}{693} +
\frac{272 \eta_n^2}{99} + \frac{6}{11\eta_n}
\,
\end{eqnarray}
We note that the leading terms in each case (and quite generally for all
values of $p$, using a recursion relation argument as above) are identical
(with $\zeta_n \leftrightarrow \eta_n$), but that the next-to-leading orders
reflect differences between the classical and quantum probability densities.
These differences vanish in the large $n$ limit.

In the context of the `half'-SHO,
we have already noted that there can be similar agreement between the
exact quantum-mechanical expectation value of $x$
in Eqn.~(\ref{large_n_quantum})
and the corresponding classical result
in Eqn.~(\ref{classical_half_sho_result}) in the large $n$ limit.
One can use the recursion relations in Eqn.~(\ref{full_sho_recursion_relation})
to obtain the highest-order terms in the expectation values of $x^p$
and one again finds agreement with the semi-classical results for large $n$.
For example, the classical result is
\begin{equation}
\langle n |x^p| n\rangle _{CL}
=
\frac{2}{\pi} \int_{0}^{A_n} \frac{x^p}{\sqrt{A_n^2 - x^2}}\,dx
= \frac{2(A_n)^p}{\pi} \int_{0}^{1} \frac{y^p\,dy}{\sqrt{1-y^2}}
= \frac{(A_n)^p}{\sqrt{\pi}} \frac{\Gamma((1+p)/2)}{\Gamma(1+p/2)}
\,.
\end{equation}

Given that traditional sum rules involve transition matrix elements,
there is no reason to expect that semi-classical probability arguments
will provide any useful information on their evaluation.
One important exception, however, is the form
of the second-order perturbation theory result for the energy,
as in Eqn.~(\ref{general_second_order_shift}), which is of the form
of an energy-difference weighted sum rule. In that special case,
semi-classical expressions for the quantized energy, such as the
WKB approximation, can sometimes provide guidance on the form of the
energies, at least in the large $n$ limit.

An example of such a connection is the use of approximate WKB-type methods
in the evaluation of first-order perturbation theory results using classical
probability densities, as in Ref.~\cite{robinett_wkb}. More surprisingly,
it has been pointed out that WKB energy quantization methods can give the
correct large $n$ behavior of the second-order energy shift due to the
Stark effect in two familiar model
systems, the harmonic oscillator and infinite well \cite{robinett_polar}.
This approach was used  in Ref.~\cite{robinett_stark} where the
exact result for the second-order energy shift due to a constant
external field for the symmetric linear potential was derived for the
first time, giving the results in Eqn.~(\ref{symmetric_linear_stark_shift}).
The WKB prediction for this case is given by the quantization condition
\begin{equation}
\sqrt{2m} \int_{A_{-}}^{A_{+}}
\sqrt{E_n - (F|z| + \overline{F}z)}\,dz = (n+1/2)\hbar \pi
\end{equation}
where $n=0,1,2,..$ and the classical turning points are
$A_{\pm} = \pm E_n/(F \pm \overline{F})$. The WKB prediction for the
energies are then
\begin{equation}
E_n =
E_n^{(0)}
\left(1 - \frac{\overline{F}^2}{F^2}\right)^{2/3}
\approx
E_n^{(0)}
\left(1 - \frac{2}{3} \left(\frac{\overline{F}}{F}\right)^2
+ \cdots \right)
\end{equation}
where $E_n = {\cal E}_0 (3\pi(n+3/4)/2)^{2/3}$ is the zero-field
WKB approximation, which agrees with the exact results in
Eqn.~(\ref{large_n_expansion}) for large $n$. This implies that the
first-order Stark shift vanishes (as it must by symmetry) and that the
second-order terms are
\begin{equation}
E_n^{(2)} = - \frac{6}{9} \left(\frac{\overline{F}}{F}\right)^2
E_n^{(0)}
\,.
\end{equation}
As pointed out in Ref.~\cite{robinett_stark}, this semi-classical
result brackets the two exact quantum mechanical expressions for
the even and odd states in Eqn.~(\ref{symmetric_linear_stark_shift}),
giving it as the `average' effect.

Prompted by this partial success, we wish to examine to what extent
a similar WKB-type analysis will give reliable answers for the
first- and second-order energy shifts due to the Stark effect for the
`half'-SHO discussed above.  We first note that the WKB result for the
energy eigenvalues for the potential in Eqn.~(\ref{half_sho}) without an
external field are given by
\begin{equation}
\sqrt{2m} \int_{0}^{A_n} \sqrt{E_n - m\omega^2x^2/2}\,dx
= (n + C_L + C_R)\hbar \pi
\end{equation}
where $C_L,C_R$ are the appropriate matching constants for an
infinite wall and `linear' potential respectively, given by
$C_L = 1/2$ and $C_R = 1/4$, respectively. Evaluating the integral,
the WKB prediction is then
given by $E_n = (2n+3/2)\hbar \omega$, reproducing the exact result.

The corresponding expression including a perturbing
linear field, $\overline{V}(x)
= \overline{F}x$, is then
\begin{equation}
\sqrt{2m} \int_{0}^{A^{(+)}_n} \sqrt{E - m\omega^2x^2/2 - \overline{F}x}\,dx
= (n + 3/4)\hbar \pi
\end{equation}
where the upper turning point is given by energy conservation to be
\begin{equation}
A^{(+)}_n = \frac{\sqrt{2m\omega^2 E_n + \overline{F}^2} - \overline{F}}
{m\omega^2}
\, .
\end{equation}
The integral can be done in closed form, but we only require the result
expanded to second order in $\overline{F}$, which gives
\begin{equation}
\frac{\pi}{2} \sqrt{\frac{m}{m\omega^2}}
\left[ E_n - \frac{2}{\pi} \sqrt{\frac{2E_n}{m\omega^2}}
\overline{F}
+ \frac{\overline{F}^2}{2m\omega^2}\right]
= (n+3/4)\hbar \pi
\, .
\end{equation}
This can be rationalized to give the simple quadratic equation
\begin{equation}
E_n = R_n \pm \sqrt{R_n^2 - Z_n^2}
\end{equation}
where
\begin{equation}
Z_n \equiv \frac{\overline{F}^2}{2m\omega^2}
\qquad
\quad
\mbox{and}
\quad
\qquad
R_n = Z_n + \frac{4\overline{F}^2}{\pi^2 m\omega^2}
\,.
\end{equation}
Solving for $E_n$, again as a series in $\overline{F}$, we find that
\begin{eqnarray}
E_n^{(0)}(WKB) & = & (2n+3/2) \hbar \omega
\label{zero_order}\\
E_n^{(1)}(WKB) & = & \frac{2\overline{F}}{\pi}
\sqrt{\frac{2E_n^{(0)}}{m\omega^2}}
\label{first_order}\\
E_n^{(2)}(WKB) & = & \left( - \frac{1}{2} + \frac{4}{\pi^2}\right)
\frac{\overline{F}^2}{m\omega^2}
\label{second_order}
\,.
\end{eqnarray}
The zero-order result is the standard WKB prediction noted above,
while the first-order expression coincides with first-order
perturbation theory (using the matrix element in Eqn.~(\ref{large_n_quantum})
in the large $n$ limit.

The second-order result is more interesting. The exact quantum mechanical
result for the second-order Stark effect for the ordinary harmonic
oscillator is $E_n^{(2)} = - \overline{F}^2/2m\omega^2$ which is most
easily obtained by a simple change of variables in the original
Schr\"{o}dinger equation, and trivially confirmed in second-order perturbation
theory, using the matrix elements in Eqn.~(\ref{full_sho_1}). For
the `half'-SHO, the WKB prediction is still a constant negative
shift, the same for all states, but with a non-trivially different
coefficient. We can find no simple way to extract the exact second-order
result from a direct solution of the differential equation, but the
second-order perturbation theory expression in
Eqn.~(\ref{general_second_order_shift}) can
be evaluated numerically using the dipole matrix elements in
Eqn.~(\ref{half_sho_dipole_matrix_element}).

\begin{figure}[hbt]
\epsfig{file=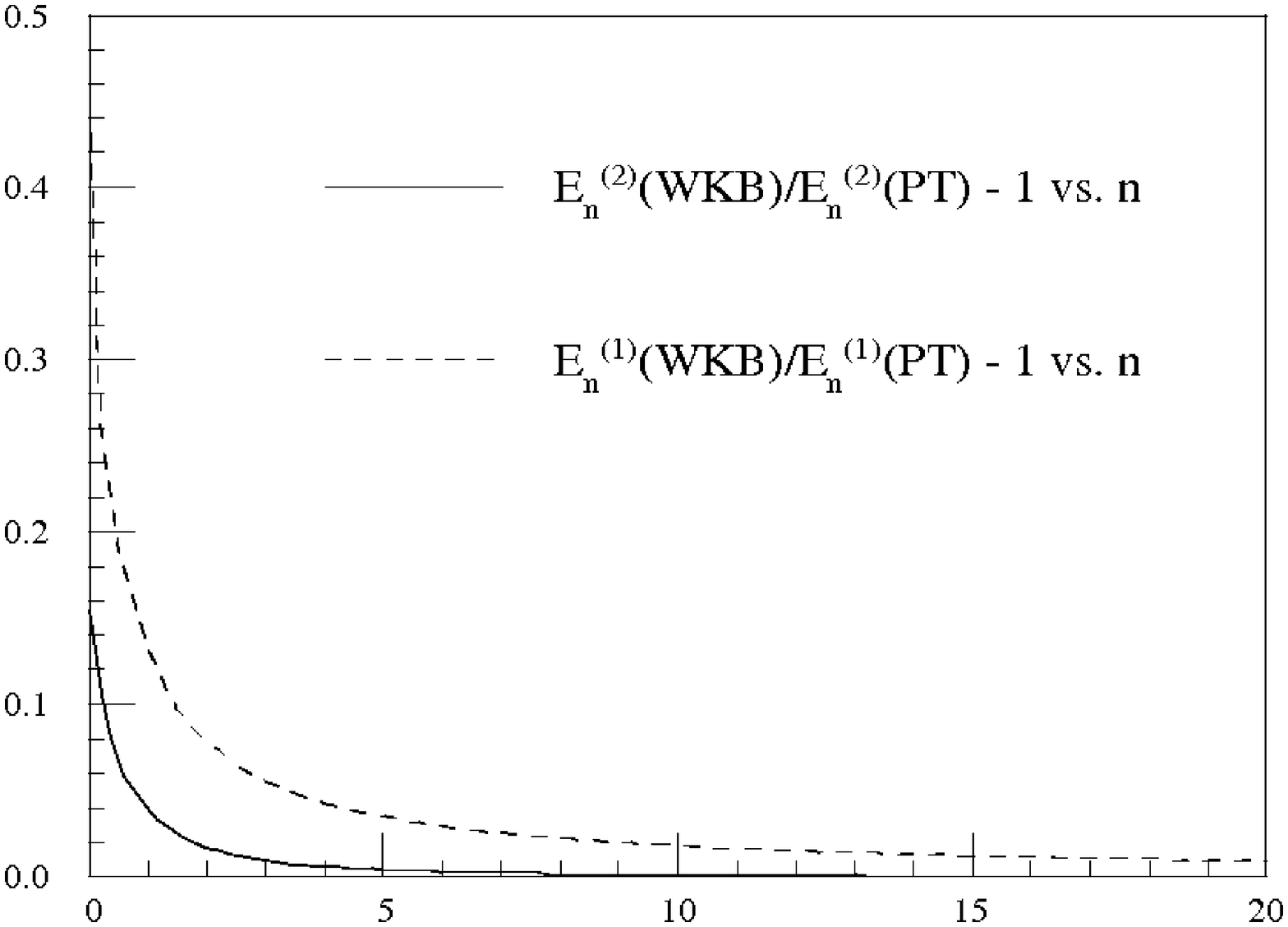,width=0.7\linewidth,angle=270}
\caption{Comparison of the WKB estimate (WKB) of the first- and second-order
energies for the Stark effect (from Eqns.~(\ref{first_order})
and (\ref{second_order}))
versus the exact results from perturbation
theory (PT). The quantity $E_n^{(1,2)}(WKB)/E_n^{(1,2)}(PT) -1$ is
shown as dashed (first-order) and dashed (second-order) respectively.}
\label{fig:classical_quantum_comparison}
\end{figure}

To compare the first- and second-order predictions from the WKB approach
with the exact results from first- and second-order perturbation theory (PT),
we plot the differences between the WKB and PT methods in
Fig.~\ref{fig:classical_quantum_comparison},
as a function of the quantum number $n$.  As mentioned above,
the first order predictions in both approaches agree in the large $n$ limit,
which we've demonstrated here analytically.
We note the more interesting result that
the simple expression in Eqn.~(\ref{second_order}) gives the correct
large $n$ behavior of the second-order Stark shifts for the `half'-SHO
problem, reproducing a non-trivial numerical factor which would otherwise
be difficult to extract.

We note that this is another example of where a WKB approach to the
evaluation of the second-order Stark effect correctly predicts the
large $n$ behavior. This further justifies the discussion
Ref.~\cite{robinett_polar} where such an approach was used to
systematically evaluate the second-order energy shifts due to an
external field for general power-law potentials, $V_{k}(x) = V_0|x/a|^k$.

\section{Conclusions and Discussion}

We have examined two cases of parity-related potentials, the quantum
bouncer extended to the symmetric linear potential, and the harmonic
oscillator reduced to the `half'-SHO, in order to probe the importance of
the continuity of the potential on the convergence of quantum mechanical
sum rules. We have indeed seen that the smoothness of $V(x)$ has
clear consequences on which sum rules will be realized, which in turn is
closely related to
the convergence properties of the individual matrix elements. For the
symmetric linear potential, we find new constraints on the zeros of the
derivative of the Airy function, but note that they are very similar in
functional form to those derived from the quantum bouncer. On the other hand,
the infinite set of constraints which arise from the `half'-SHO are
qualitatively very different from the super-convergent sums found
in the realization of sum rules for the more familiar oscillator.

The study of parity-related potentials is also motivated by the desire to
find examples where there is a substantial overlap between the energies
and wavefunctions (needed in the evaluation of matrix elements)
of two quantum systems as they relate to sum rules.
 That connection is realized here by the fact that
the odd states of a symmetric potential remain solutions of the
parity-restricted version, so that all of the resulting energies and
wavefunctions (save a trivial normalization) are still solutions of the
parity-restricted partner potential.

Another class of quantum mechanical problems which has similar strong
connections between the energy levels and wavefunctions are
super-partner potentials \cite{sukhatme}, $V^{(-)}(x)$ and
$V^{(+)}(x)$, in the context of supersymmetric quantum mechanics (SUSY-QM).
 In that case, the spectra of the two systems are identical,
except for the zero-energy ground state ($E_0^{(-)} = 0$) of the first
system, which is absent in the second.
 Another reason for us to consider the `half'-SHO potential in such detail
is that it is easily extended to generate an appropriate $V^{(\pm)}(x)$ pair
in SUSY-QM.  For example, the potential
\begin{equation}
V^{(-)}(x) = \frac{1}{2}m\omega^2 x^2 - \frac{3}{2}\hbar \omega
\qquad
\mbox{for $x\geq 0$}
\end{equation}
has the energy spectrum $E_n^{(-)} = 2n\hbar \omega$ for $n=0,1,2,...$
with the ground state wavefunction
\begin{equation}
\psi_0(x) = \frac{2x}{\sqrt{\beta^3\sqrt{\pi}}}
\, e^{-(x/\beta)^2/2}
\qquad
\mbox{for $x\geq0$}
\end{equation}
and the remaining states given by Eqn~(\ref{half_sho_solutions}).

Using this ground-state solution to form the super-potential, we find
\begin{equation}
W(x) =
-\frac{\hbar}{\sqrt{2m}}
\left( \frac{\psi_0'(x)}{\psi_0(x)}\right)
=
-\frac{\hbar}{\sqrt{2m}}
\left( \frac{1}{x} - \frac{x}{\beta^2}\right)
\end{equation}
allowing us to construct the super-partner potential
\begin{equation}
V^{(+)}(x) = \frac{1}{2} m\omega^2 x^2 + \frac{2\hbar^2}{2mx^2}
- \frac{1}{2} \hbar \omega
\,.
\end{equation}
This has the form of the radial equation for the three-dimensional harmonic
oscillator with the special choice of the angular momentum quantum
 number $l=1$ (giving the $l(l+1) = 2$ factor in the centrifugal barrier term)
and an overall constant shift in energy. Using standard
results for the energy eigenvalues for that system, we find that
\begin{equation}
E_n^{(+)} = \hbar \omega \left(2n +l + \frac{3}{2}\right) - \frac{1}{2}\hbar
\omega
= 2\hbar \omega (n+1)
\end{equation}
for the relevant $l=1$ case. One can also use standard textbook results
to obtain the properly normalized solutions to be
\begin{equation}
\psi_n^{(+)}(x) = N_n x^2\, e^{-(x/\beta)^2/2}\, L_n^{(3/2)}(x^2/\beta^2)
\qquad
\mbox{with}
\qquad
N_n = \sqrt{\frac{2^{k+3} k!}{\beta^5 (2k+3)!! \sqrt{\pi}}}
\,.
\end{equation}
This is one example of many familiar super-partner potentials
\cite{sukhatme} which can
be systematically studied in the context of quantum mechanics
to probe the delicate interplay between energy level differences and
matrix elements which must exist to guarantee the realization of the
infinite number of sum rules which one can generate using the simple
procedures outlined here.

\section{Acknowledgments}
O.A.A., K.C., and M.B. were funded in part by a Davidson College Faculty Study and 
Research Grant and by the National Science Foundation (DUE-0442581).

\appendix

\section{Necessary Integrals Involving Airy function}
\label{sec:indefinite}

In this Appendix, we collect indefinite integrals of products of
Airy functions, useful for normalizations, expectation values, and
matrix elements  as first noted by Gordon \cite{gordon}
and Albright \cite{albright}, adding
one new result (Eqn.~(\ref{off_diagonal_2}).

We then extend the recursion relation for general matrix elements of
Airy functions, first derived by Goodmanson \cite{goodmanson},
necessary for both the even and odd states of the symmetric linear
potential in Sec.~\ref{sec:recursion} and also
generalize those recursion relations to the
case of the harmonic oscillator problem, allowing one to evaluate
matrix elements for the `half'-SHO problem.

We first assume two arbitrary linear combinations of the Airy differential
equation,
\begin{equation}
A(\zeta-\beta)  =  a Ai(\zeta-\beta) + b Bi(\zeta-\beta)
\qquad
\mbox{and}
\qquad
B(\zeta-\beta)  =  c Ai(\zeta-\beta) + d Bi(\zeta-\beta)
\end{equation}
where
\begin{equation}
A''(\zeta-\beta) = (\zeta-\beta) A(\zeta-\beta)
\qquad
\quad
\mbox{and}
\quad
\qquad
B''(\zeta-\beta) = (\zeta-\beta) B(\zeta-\beta)
\, .
\end{equation}
In the expressions below, when we refer to integrals of such
solutions corresponding to different values of the `shift', $\beta_1
\neq \beta_2$, we write $A_1$ for $A(\zeta-\beta_1)$ and $B_2$
for $B(\zeta-\beta_2$).

Each of the indefinite integrals
we need can be written in terms of four basic quantities, namely
\begin{eqnarray}
F_1 & = & \{AB\} \label{term_1}\\
F_2 & = & \{A'B - AB'\} \label{term_2}\\
F_3 & = & \{A'B + AB'\} \label{term_3}\\
F_4 & = & \{A'B'\} \label{term_4}
\end{eqnarray}
times polynomials in $\zeta$ and the appropriate values of $\beta$.

For the integrals involving solutions with the same `shift', we
have
\begin{eqnarray}
\!\!\!\!
\int \, A\, B\, d\zeta
&  =  &  (\zeta-\beta) \, F_1 - F_4
\label{diagonal_0} \\
\int \, \zeta\, A\, B\, d\zeta
& = &
\frac{1}{3} (\zeta^2 + \beta \zeta - 2\beta^2) \,F_1
+\frac{1}{6}\,F_3
- \frac{1}{3} (\zeta+2\beta)\, F_4 \label{diagonal_1}\\
 \int \, \zeta^2\, A \, B\, d\zeta
& = &
\frac{1}{15} (3\zeta^3 + \beta \zeta^2 + 4\beta^2 \zeta - 8\beta^3  -3)
\, F_1
+ \frac{1}{15}(3c + 2\beta) \, F_3
\nonumber \\
& &
\qquad
- \frac{1}{15}(3\zeta^2 + 4\beta \zeta + 8\beta^2) F_4
\label{diagonal_2}
\, .
\end{eqnarray}
For integrals involving derivatives, we can modify one of the
identities in Albright \cite{albright} to write for shifted
solutions
\begin{eqnarray}
\int A' \, B'\, d\zeta
& = & -\frac{1}{3}(\zeta-\beta)^2 \,F_1
+ \frac{1}{3}\, F_3
+ \frac{1}{3}(\zeta-\beta)\, F_4
\, .
\label{diagonal_derivative}
\end{eqnarray}

Finally, for integrals with two different shifts ($\beta_1 \neq \beta_2$),
we have results for $p=0,1$ from Gordon \cite{gordon}
and a new result derived here
for $p=2$. Recall that the $T_{i}$ below will be constructed
from the terms in Eqns.~(\ref{term_1}) - (\ref{term_4}) using
$A_1 = A(\zeta-\beta_1)$  and $B_2 = B(\zeta-\beta_2)$.
\begin{eqnarray}
\!\!\!\!\!\!\!
\int \, A_{1} \, B_{2}\, d\zeta
& = & \frac{1}{(\beta_2-\beta_1)}
\, F_2
\label{off_diagonal_0} \\
\int \, \zeta\, A_{1} \, B_{2}\, d\zeta
& = &
\frac{(\beta_1 + \beta_2 - 2\zeta)}{(\beta_1 - \beta_2)^2}
\, F_1
+
\left\{
\frac{\zeta}{(\beta_2 - \beta_1)}
+
\frac{2}{(\beta_2 - \beta_1)^3}
\right\}
\,F_2 \nonumber \\
& &
\qquad \qquad + \frac{2}{(\beta_1-\beta_2)^2}
\, F_4
\label{off_diagonal_1} \\
\int \, \zeta^2\, A_{1} \, B_{2}\, d\zeta
& = &
\left[ \frac{12(\beta_1+\beta_2)}{(\beta_1-\beta_2)^4}
+ \frac{2(-12 + (\beta_1+\beta_2)(\beta_1-\beta_2)^2)}{(\beta_1-\beta_2)^4}\zeta
- \frac{4}{(\beta_1- \beta_2)^2} \zeta^2\right]\, F_1
\nonumber \\
& &
+ \left[\frac{4(-6 +(\beta_1+\beta_1)(\beta_1 - \beta_2)^2)}{(\beta_1 - \beta_2)^5}
- \frac{12}{(\beta_1 - \beta_2)^3} \zeta
- \frac{1}{(\beta_1 - \beta_2)} \zeta^2\right]\, F_2
\nonumber \\
& & - \left[\frac{2}{(\beta_1-\beta_2)^2}\right]\, F_3
 + \left[\frac{24}{(\beta_1-\beta_2)^4} +
\frac{4}{(\beta_1-\beta_2)^2} \zeta\right]
\, F_4
\label{off_diagonal_2}
\end{eqnarray}

\section{Recursion relation for matrix elements of Airy functions and
harmonic oscillator solutions}
\label{sec:recursion}

Goodmanson \cite{goodmanson} derived a recursion relation relating the
power-law matrix elements of solutions for the quantum bouncer problem,
involving integrals of the form
\begin{equation}
R_p = \int_{0}^{\infty} \zeta^p\,Ai(\zeta-\zeta_n)\, Ai(\zeta-\zeta_m)\,d\zeta
\end{equation}
assuming that the solutions satisfied the boundary condition for the
bouncer problem, namely that $Ai(-\zeta_{n,m}) = 0$. One can easily repeat
his analysis, not making that assumption,
to find a more general recursion relation
valid for integrals relevant for both the even and odd states of the
symmetric linear potential. We can also repeat the analysis to obtain
relations on matrix elements of the solutions of the harmonic oscillator,
which will prove useful for the parity-restricted `half'-SHO in
Sec.~\ref{sec:half_sho}.

\subsubsection{Recursion relations for Airy functions}
\label{subsubsec:airy}

We begin by adopting the notation
\begin{equation}
A_{n} = A_{n}(\zeta) \equiv Ai(\zeta-\zeta_{n})
\qquad
\mbox{where}
\qquad
A''_{n} = A''_{n}(\zeta) = (\zeta-\zeta_{n}) A_{n}
\end{equation}
and then assume a general $f(\zeta)$ as a well-behaved function, at least multiply
differentiable and
satisfying $\lim_{\zeta \rightarrow \infty} f(\zeta)Ai[\zeta]^2 = 0$. We will, in fact,
use $f(\zeta) = \zeta^p$ to evaluate various matrix elements.

We start with the mathematical identity
\begin{equation}
\int_{0}^{\infty}
\left[2 f'(\zeta) A_n' A_m' - f''(\zeta)(A_n A_m)'\right]'\,d\zeta
=
\left[2 f'(\zeta) A_n' A_m' - f''(\zeta)(A_n A_m)'\right]_{0}^{\infty}
\end{equation}
and then use integration by parts multiple times to isolate the
$A_n(\zeta) A_m(\zeta)$ terms times derivatives of $f(\zeta)$ (on the left hand side)
and resulting surface terms (on the right hand side.). The
resulting integral can be written in the form
\begin{equation}
I
\equiv
\int_{0}^{\infty}
A_n\,A_m\,
\left[ f^{(iv)}(\zeta) - 4(\zeta-\zeta_{ave})f''(\zeta)  - 2f'(\zeta) + (\zeta_{n}-\zeta_{m})^2 f(\zeta)
\right] d\zeta
\end{equation}
which can be shown to be equal to
\begin{eqnarray}
I & = &
 \left[2 f'(\zeta) A_n' A_m' - f''(\zeta)(A_n A_m)'\right]_{0}^{\infty}
+ \left[f'''(\zeta)A_n A_m\right]_{0}^{\infty}
- \left[2A_n A_m (\zeta-\zeta_{ave})f'(\zeta)\right]_{0}^{\infty}
\nonumber \\
& &
\qquad \qquad
- \left[(\zeta_{n}-\zeta_{m}) (A'_nA_m - A_nA'_m)f(\zeta)\right]_{0}^{\infty}
\nonumber \\
& = &
-2f'(0) A_n'(0)A_m'(0) + f''(0)
\left[A_n'(0) A_m(0) + A_n(0) A_m'(0)\right]
\nonumber \\
& &
\qquad \qquad
-f'''(0)A_n(0) A_m(0) - 2A_n(0) A_m(0) f'(0) \zeta_{ave} \nonumber \\
& &
\qquad \qquad
+ (\zeta_n-\zeta_m)f(0)
\left[A_n'(0) A_m(0) - A_n(0) A_m'(0)
\right]
\end{eqnarray}
where $\zeta_{ave} \equiv (\zeta_n+\zeta_m)/2$.
The first term on the right-hand-side reproduces the Goodmanson
\cite{goodmanson} result.
We have assumed that $f(\zeta)$ is well-enough behaved that surface terms at
infinity vanish.
One can then evaluate matrix elements over the range $(0,\infty)$
recursively for any combination of products of either even or odd
solutions, where simplifications occur since $A_{n,m}'(0) = 0$ or
$A_{n,m}(0) = 0$ respectively.

\subsubsection{Recursion relations for oscillator solutions}
\label{subsubsec:sho}

For the ordinary harmonic oscillator, expectation values and matrix
elements will be evaluated by integrals of solutions over the range
$(-\infty,+\infty)$. But an almost identical derivation to that above
allows for the evaluation of related quantities over the half-line
($0,+\infty)$ which will be important for the parity-restricted oscillator
(the `half'-SHO). For example, if we label $\psi_n(y)$ as a solution
to the dimensionless oscillator problem in  Eqn.~(\ref{dimensionless_sho}),
namely
\begin{equation}
\psi_n''(y) = (y^2 - \epsilon_n) \psi_n
\, ,
\end{equation}
we can make repeated use of that relation and integrations by part to
write
\begin{equation}
J \equiv
\int_{0}^{\infty}
\psi_n(y)\,\psi_m(y)\,
\left[ f^{(iv)}(y) - 4(y^2-\epsilon_{ave})f''(y)  - 4 yf'(y)
 + (\epsilon_{n}-\epsilon_{m})^2 f(y) \right] dy
\end{equation}
in terms of surface terms. In this way, we find that
\begin{eqnarray}
J & = & -2 f'(0) \psi_n'(0) \psi_m'(0) + \psi_n(0) \psi_m(0)
\left[ f'''(o) + 2f'(0) \epsilon_{ave}\right] \nonumber \\
& &
\qquad
f''(0) \left[ \psi_n'(0) \psi_m(0) + \psi_n(0) \psi_m'(0)\right] \nonumber \\
& &
\qquad
-(\epsilon_n -\epsilon_m) f(0)
\left[ \psi_n'(0) \psi_m(0) - \psi_n(0) \psi_m'(0)\right]
\label{full_sho_recursion_relation}
\end{eqnarray}
where $\epsilon_{ave} = (\epsilon_n + \epsilon_m)/2$.
Note that for the `half'-SHO, only the first term where $\psi_n'(0)$
and $\psi_m'(0)$ are non-zero

We can confirm that this expression encodes the absolute selection rules
for matrix elements of the standard oscillator,
as in Eqns.~(\ref{full_sho_1}) - (\ref{full_sho_3}). For example,
for the dipole matrix elements, we know that for the oscillator only
even-odd matrix elements will be relevant, so using $f(y) = y$, we find
that
\begin{equation}
J = [(\epsilon_n - \epsilon_m)^2 -4]\int_{0}^{\infty}
\psi_n(y)\, \psi_m(y)\,dy = 0
\end{equation}
so that the dipole matrix elements vanish unless $(n-m)^2 = 1$
or $n = m \pm 1$.

\section{Semi-classical probability density for the quantum
bouncer}
\label{sec:semiclassical_limit}

It is straightforward to see how the classical probability distribution
in Eqn.~(\ref{classical_bouncer_distribution})
for the quantum bouncer arises as the locally averaged
limit of the exact quantum mechanical solution. For the bouncer, the solution
is given by
\begin{equation}
\psi_n(z) = \sqrt{2} N_n^{(-)} Ai\left(\frac{z}{\rho} - \zeta_n\right)
\end{equation}
where $N_n^{(-)}$ is given by Eqn.~(\ref{normalizations}).
Standard handbook \cite{stegun} results exist for the behavior of
$Ai(\zeta)$ and $Ai'(\zeta)$ for large negative values of $\zeta$, namely
\begin{eqnarray}
Ai(-\zeta) & \sim & \pi^{-1/2} \zeta^{-1/4}
\sin\left(\xi + \frac{\pi}{4}\right) \\
Ai'(-\zeta) & \sim & \pi^{-1/2} \zeta^{+1/4}
\cos\left(\xi + \frac{\pi}{4}\right)
\end{eqnarray}
where $\xi \equiv 2 x^{3/2}/3$.
The quantum-mechanical probability density is then given by
\begin{eqnarray}
P_{QM}^{(n)}(z) = |\psi_n(z)|^2
& = &
\frac{1}{\rho} \left[\frac{Ai(z/\rho -\zeta_n)}{Ai'(-\zeta_n)}\right]^2
\\
& \sim &
\frac{\sin^2\left[2(\zeta_n-z/\rho)^{3/2}/3 + \pi/4\right]}
{\rho \sqrt{\zeta_n (\zeta_n - z/\rho)}}
\end{eqnarray}
where we use use the expression for $\zeta_n$ in the large $n$ limit
in Eqn.~(\ref{large_n_expansion}) to write
\begin{equation}
\cos\left( \frac{2}{3}\xi + \frac{\pi}{4}\right)
\sim
\cos\left\{ \frac{2}{3}\left( \left[\frac{3\pi}{2}(n-1/4)\right]^{2/3}
\right)^{3/2} + \frac{\pi}{4}\right\}
= \cos(n\pi) = \pm 1
\end{equation}
In the semi-classical limit, the oscillatory component locally averages
to $1/2$, giving the purely classical result in
Eqn.~(\ref{purely_classical_bouncer}). This explicit derivation agrees
with the standard result for a purely classical probability distribution
for position, namely
\begin{equation}
P_{CL}(x) = \frac{2}{\tau} \sqrt{\frac{m}{2(E-V(x))}}
\end{equation}
where $\tau$ is the classical period,
given by
\begin{equation}
\frac{\tau}{2}
= \int_{a}^{b}\frac{dx}{v(x)}
\end{equation}
with $v(x)$ being the classical speed and $a,b$ are the classical turning
points for bounded motion.

\newpage

\noindent
\hfill

\hfill

\end{document}